\documentclass[aps,prl,twocolumn,floatfix]{revtex4}

\usepackage{epsfig}

\def\theta{\vartheta}
\def\dcp{\delta_{\rm CP}}

\usepackage[dvips]{color}
\definecolor{Black}{named}{Black}
\definecolor{Red}{named}{Red}

\begin{document}

\title{Measuring the 13-mixing angle and the CP phase with neutrino
  telescopes}

\author{P.~D.~Serpico and M.~Kachelrie\ss}

\affiliation{Max-Planck-Institut f\"ur Physik
(Werner-Heisenberg-Institut), F\"ohringer Ring 6, 80805 M\"unchen, Germany}

\date{April 7, 2005 -- version 2}

\begin{abstract}
The observed excess of high-energy cosmic rays from the Galactic plane
in the energy range around $10^{18}$~eV may be naturally explained by
neutron primaries generated in the photo-dissociation of heavy nuclei.
In this scenario, lower-energy neutrons decay
before reaching the Earth and produce a detectable flux in a 1~km$^3$
neutrino telescope. The initial flavor composition of the neutrino flux,
$\phi(\bar\nu_e):\phi(\bar\nu_\mu):\phi(\bar\nu_\tau)=1:0:0$,
offers the opportunity to perform a combined $\bar\nu_\mu/\bar\nu_\tau$
appearance and $\bar\nu_e$ disappearance experiment.
The observable flux ratio $\phi(\bar\nu_\mu)/\phi(\bar\nu_e+\bar\nu_\tau)$
arriving at Earth depends appreciably on the 13-mixing angle
$\theta_{13}$ and the leptonic CP phase $\dcp$, thus opening a new
experimental avenue to measure these two quantities.
\end{abstract}

\pacs{
14.60.Pq, 
95.85.Ry,
98.70.Sa    
\hfill Preprint MPP-2005-10
}

\maketitle

{\em Introduction.---}%
Neutrino physics has progressed enormously in the last decade.
The discovery of neutrino oscillations provides the first clear
experimental signature for the incompleteness of the Standard Model of
particle physics~\cite{review}. While the determination of the
mixing parameters controlling the solar and atmospheric neutrino
oscillations has already entered the precision era, it
exists currently only an upper limit for the 13-mixing angle
$\theta_{13}$ from the CHOOZ experiment, 
$\sin^2 2\theta_{13}<0.1$~\cite{Apollonio:2002gd}.  
The mixing angle $\theta_{13}$
characterizes how strong atmospheric and solar oscillations
are coupled and therefore also determines the strength of CP violation
effects in neutrino oscillations. Among the three possible
phases in the neutrino mixing matrix, only the Dirac phase $\dcp$
enters neutrino oscillations. This phase is at present
completely unconstrained.
Both the mixing angle $\theta_{13}$ and the phase $\dcp$ are observable
in solar and atmospheric neutrino oscillation experiments only as
subleading, genuine three-flavor effects that are masked
mainly by systematic uncertainties~\cite{sub}.
While there are strong experimental efforts to improve the
measurements of $\theta_{13}$ in the near future by dedicated 
experiments~\cite{f1}, the detection 
of a non-zero $\dcp$ appears unlikely for the next generation
of facilities~\cite{Huber:2004ug}. 
Thus the answer to one of the most interesting questions in neutrino 
physics, namely the existence of leptonic CP violation, probably has
 to await the construction of
long-baseline experiments using second-generation superbeams or
perhaps even a neutrino factory.

In the following, we propose to use high-energy neutrinos produced by
decaying neutrons as a new probe to measure $\theta_{13}$ and $\dcp$
with neutrino telescopes.
The potential of neutrino telescopes to measure the atmospheric mixing
angle $\theta_{12}$  has been discussed recently in Ref.~\cite{BG},
while tests for new physics beyond standard neutrino oscillations
using atmospheric neutrino data have been examined in Ref.~\cite{GG}.
Previously, Refs.~\cite{Beacom:2002vi,Beacom:2003zg} discussed
possibilities to measure or to constrain $\theta_{13}$ and $\dcp$ for
the case of decaying neutrinos, while the use of the neutron decay
channel as neutrino source was suggested in Ref.~\cite{Hooper:2004xr}
to test quantum decoherence. Neutron primaries have been invoked to
explain an excess of high-energy cosmic rays (CRs) from two regions
in the Galactic plane~\cite{Anchordoqui:2003vc,Crocker:2004}. 
This signal, in a limited energy range around
$10^{18}$~eV, has been observed by several experiments with different
techniques:
The AGASA collaboration found a correlation of the arrival directions of
CRs with the Galactic plane at the 4$\:\sigma$
level~\cite{Hayashida:1998qb}. This excess, which is roughly 4\% of
the diffuse flux, is concentrated towards the Cygnus region, with a
second hot spot towards the Galactic Center (GC)~\cite{Teshima2001}.
Such a signal has been independently confirmed by the Fly's Eye
Collaboration~\cite{Bird:1998nu} and by a re-analysis of the SUGAR
data~\cite{Bellido:2000tr}.

Complementary evidence for a cosmic accelerator in the Cygnus region 
comes from the detection of an extended TeV~$\gamma$-ray source by the
HEGRA experiment~\cite{Aharonian:2002ij,Aharonian:2005}. The
measured photon spectrum is difficult to explain in terms of electromagnetic
acceleration. Also, X-ray or radiowave emission could not be
detected by CHANDRA or VLA~\cite{Butt:2003xc}, thus favoring
a hadronic accelerator.
Similarly, multi-TeV $\gamma$-rays from the vicinity of the GC have been
recently detected by HESS~\cite{Aharonian:2004wa}.

{\em Galactic neutron sources.---}%
The excess from the Cygnus and GC region is seen at  $E\approx 10^{18}$~eV,
i.e. at energies where charged cosmic rays still suffer large deflections in
the Galactic magnetic field so that only a neutral primary can produce a
directional signal. Another evidence for neutrons as primaries is that
the signal appears just at that energy where the neutron lifetime
allows neutrons to propagate from a distance of several kpc.

Neutrons can be generated as secondaries either in collisions of
high-energy protons on ambient photons and protons, or in the
photo-dissociation of heavy nuclei. In the first case, the flux of
$\bar\nu_e$ from neutron decays would be negligible compared
to the neutrino flux from pion decays. Thus one expects a neutrino
flavor composition of
$\phi_e:\phi_\mu:\phi_\tau=1:2:0$ before oscillations~\footnote{We 
denote with $\phi_\alpha$ the combined flux of
  $\nu_\alpha$ and $\bar\nu_\alpha$.}, 
typical for most sources of
high-energy neutrinos. The oscillation phenomenology and signature 
for such a ``standard'' GC source were already considered 
in~\cite{Crocker:1999yw}. In contrast, photo-dissociation of heavy
nuclei produces a pure $\bar\nu_e$ initial flux. Since the energy
fraction transferred to the $\bar\nu_e$ is typically $\sim 10^{-3}$  
and only neutrons with $E\alt 10^{18}$~eV can decay on galactic
distances, the neutrino flux from  photo-dissociation is limited to
sub-PeV energies. Moreover, the threshold for photo-dissociation
on UV photons implies a lower cut-off at $E\sim$~TeV  for the
$\bar\nu_e$ energies.

There are several arguments in favor of the dominance of heavy nuclei
in the diffuse Galactic CR flux at $E\sim 10^{18}$~eV. 
First, the end of the Galactic CR spectrum is expected to consist 
of heavy nuclei, because the Galactic magnetic field confines more easily CRs
with small rigidity. Subtracting the spectrum expected for
extragalactic CRs from the measured CR spectrum,
Ref.~\cite{Berezinsky:2004wx} found evidence that the transition
between Galactic and extragalactic CRs happens around a 
${\rm few}\times 10^{17}$~eV. In this case, the total diffuse CR flux between
$(1-10)\times 10^{17}$~eV consists only of galactic iron nuclei and
extragalactic protons. 
Another method to determine the transition
energy is to study the chemical composition of the CR
flux~\cite{watson}. 
At present, these measurements are not fully conclusive but point to a
dominantly heavy component in the CR flux at least up to $\sim
10^{18}$~eV and a possible transition to extragalactic protons at
higher energies. Such a higher transition energy would also ease the
difficult luminosity 
requirements needed for extragalactic ultra-high energy cosmic ray
sources~\cite{ss}. Around and above the transition energy, the
unconfined flux from Galactic point sources becomes
visible. If the flux from these point sources consists of protons
or nuclei, has to be answered experimentally for each source
separately.

In the following we use as our basic assumption that
photo-dissociation of heavy nuclei is the origin of the decaying
neutrons. We assume first that other neutrino sources that
contaminate the pure $\bar\nu_e$ initial flux can be neglected, but
at the end we discuss how our conclusions change when this 
assumption (that can be verified experimentally) is relaxed. 
To be specific, we use the model of Anchordoqui {\it et al.} in 
Ref.~\cite{Anchordoqui:2003vc}, who calculated the neutrino flux 
from the Cygnus region which is in the field of view of the 
km$^3$ telescope ICECUBE~\cite{ice}.
These authors estimated an integrated $\bar{\nu}_e$ flux from
neutron decays of $\sim 2 \times 10^{-11} {\rm cm}^{-2} {\rm
s}^{-1}$ at $E>1$~TeV by normalizing the neutron flux to the 4\%
anisotropic component observed by AGASA. This flux corresponds to
$\approx 20$ events (of all flavors) per year in ICECUBE.

{\em Flavor composition after oscillations.---}%
%
%
The fluxes $\phi_\beta^D$ arriving at the detector are given
in terms of the probabilities $P_{\alpha\beta}\equiv
P(\bar{\nu}_\alpha\to\bar{\nu}_\beta)$~\footnote{In general (see 
e.g.~\cite{Akhmedov:2004ny}) $P(\bar{\nu}_\alpha\to\bar{\nu}_\beta,\dcp,V)=
P({\nu}_\alpha\to{\nu}_\beta,-\dcp,-V)$,
where $V$ is the matter potential. Since matter effects are negligible
and the interference terms sensitive to the sign of $\dcp$ average
out, we can use equivalently $P_{\alpha\beta}$ or $P_{\bar\alpha
  \bar\beta}$.} by 
\begin{equation}
 \phi^D_\beta = \sum_\alpha P_{\alpha\beta} \phi_\alpha =
  P_{e\beta}\, \phi_e\,,
\end{equation}
where we have inserted $\phi_\alpha=(\phi_e,0,0)$. Since the galactic
distances far exceed the experimentally known oscillation lengths
even at PeV energies, the interference terms sensitive 
to the mass splittings $\Delta m^2$'s in the usual oscillation 
formula average-out. Then we can write
\begin{equation}
  P_{e\beta}=\delta_{e \beta}-2\sum_{j>k}{\rm Re}(U_{\beta j}^{*}
  U_{\beta k} U_{e j} U_{e k}^{*}) \,,\label{exacteq}
\end{equation}
where $U$ is the neutrino mixing
matrix and greek (latin) letters are used as flavor (mass) indices.

To obtain a feeling for the dependence of the fluxes on $\theta_{13}$
and $\dcp$, we give  an expansion of $P_{e \beta}$ up to second
order in $\theta_{13}$ where we use
$\theta_{12}=\frac{\pi}{6}$ and $\theta_{23}=\frac{\pi}{4}$,
\begin{eqnarray}
 P_{e e} &\approx &\frac{5}{8}-\frac{5}{4}\theta_{13}^2\,
\nonumber\\
 P_{e\mu} & \approx &\
\frac{3}{16}+\frac{\sqrt{3}}{8}\theta_{13}\cos\dcp+\frac{5\theta_{13}^2}{8}\,
\nonumber\\
 P_{e\tau} & \approx &\
\frac{3}{16}-\frac{\sqrt{3}}{8}\theta_{13}\cos\dcp+\frac{5\theta_{13}^2}{8}
\,.\label{approxeq}
\end{eqnarray}
As expected, the survival probability $P_{ee}$ (or equivalently
$\phi^D_e$) does not depend on $\dcp$ and the unitarity relation 
$\sum_{\beta} P_{e \beta}=1$ holds at each order in $\theta_{13}$. 
Moreover, the $\bar\nu_\mu$ and
$\bar\nu_\tau$ fluxes depend on $\dcp$ only via the quantity $\cos\dcp$.
Note that the independence of $P_{ee}$ from $\theta_{23}$ and $\dcp$,
as well as the relation $P_{e\mu}=P_{e\tau}(\theta_{23}\to\theta_{23}+\pi/2)$
(which shows up in the opposite signs of the $\cos\dcp$ terms in
Eq.~(\ref{approxeq})) hold exactly~\cite{Akhmedov:2004ny}.
Though the approximate relations Eq.~(\ref{approxeq}) are useful to
grasp the main features of the dependence of the fluxes $\phi^D_\alpha$
on $\theta_{13}$ and $\dcp$, in the following we will use
the exact expressions given in Eq.~(\ref{exacteq}). 
For all numerical examples, we fix the value of the solar mixing angle
to $\theta_{12}=32.5^\circ$~\cite{review}.

{\em Flavor discrimination in ICECUBE.---}%
Let us now recall briefly the flavor-discrimination possibilities
in ICECUBE~\cite{Beacom:2003nh}.
For the energies relevant here, 
$10^{12}~{\rm eV}\alt E \alt 10^{15}$~eV, the
charged-current
interactions of $\nu_e$ and $\nu_\tau$ are in principle only
distinguishable by the different muon content in electromagnetic and
hadronic showers. In practice, this is an experimental challenge and
we consider  $\nu_e$ and $\nu_\tau$ as indistinguishable in a
neutrino telescope.
By contrast, in $\nu_\mu$ charged-current interactions the long range
of muons ensures that the muon track is always visible and allows the
identification of these events. Finally, all flavors undergo the same,
indistinguishable neutral-current (NC) interactions. This interaction
contributes however only 20\% to the total cross
section~\cite{Gandhi:1998ri}. Moreover, in
this case the energy of the primary is underestimated by a factor
3--4, further suppressing the relative importance of NC interactions
because of the steeply falling energy spectrum. In the following, we
neglect therefore NC interaction and consider the combined
$\bar\nu_e$ and $\bar\nu_\tau$ flux $\phi^D_e+\phi^D_\tau$ and the
$\bar\nu_\mu$ flux $\phi^D_\mu$ as our two observables.

\begin{figure}
\epsfig{file=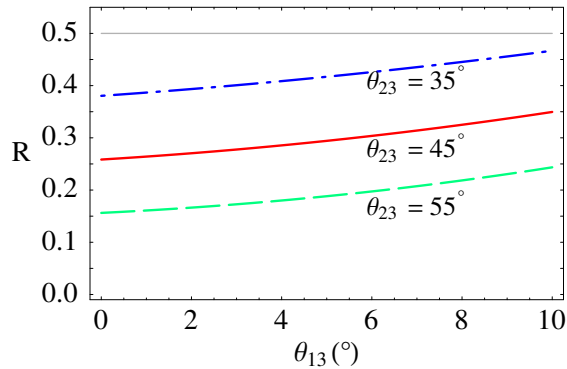,width=1.00\columnwidth}
\caption{Flux ratio $R=\phi^D_\mu/(\phi^D_e+\phi^D_\tau)$  at Earth
as a function of $\theta_{13}$ for $\theta_{23}=35^\circ$ 
(blue, dot-dashed curve), $\theta_{23}=45^\circ$ (red, solid curve),
$\theta_{23}=55^\circ$ (green, dashed curve);
for initial fluxes $\phi_e:\phi_\mu:\phi_\tau=1:0:0$ at the source and
$\dcp=0$. The ratio $R=0.5$ expected for standard astrophysical sources is 
shown for comparison.
\label{theta}}
\end{figure}

The flux ratio $R=\phi^D_\mu/(\phi^D_e+\phi^D_\tau)$ as only
observable does not allow the simultaneous measurement of $\theta_{13}$
and $\dcp$. For the sake of clarity, we first explore 
the sensitivity of $R$ to the value of $\theta_{13}$, fixing
$\dcp=0$. In Fig.~\ref{theta}, we show the expected ratio
$R$ as a function of $\theta_{13}$ for three representative values of
$\theta_{23}$. This ratio  varies by $\sim$ 50\% in the interval
$0^\circ\leq\theta_{13}\leq 10^\circ$ and differs in the extreme by a
factor of three from the standard value, 
$\phi^D_\mu/(\phi^D_e+\phi^D_\tau)=1/2$, also shown for comparison.

\begin{figure}
\epsfig{file=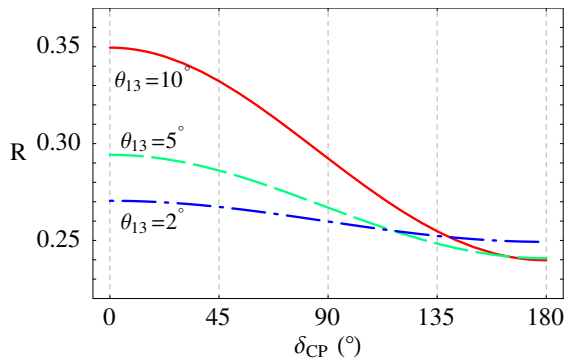,width=0.88\columnwidth}
\caption{Flux ratio $R=\phi_\mu^D/(\phi^D_e+\phi^D_\tau)$  at Earth
as a function of $\dcp$ for $\theta_{13}=10^\circ$ (red, solid curve),
$\theta_{13}=5^\circ$ (green, dashed curve), and
$\theta_{13}=2^\circ$ (blue, dot-dashed curve);
for $\theta_{23}=45^\circ$ and initial fluxes
$\phi_e:\phi_\mu:\phi_\tau=1:0:0$ at the source.
\label{dcp}}
\end{figure}

If the next generation of oscillation experiments measures or strongly 
constrains $\theta_{13}$, a neutrino telescope may even aim to detect
leptonic CP violation. In Fig.~\ref{dcp}, we show the expected ratio
$R=\phi^D_\mu/(\phi^D_e+\phi^D_\tau)$ as a function of $\dcp$ for
three values of $\theta_{13}$; we have chosen the best fit 
value $\theta_{23}=45^\circ$.
In this case the ratio varies maximally by about 40\% in 
the interval $0^\circ\leq\dcp\leq 180^\circ$ and differs in the extreme 
by a factor two from the standard value 1/2. If we use instead
$\theta_{23}=35^\circ$ ($55^\circ$),  
the only change would be an overall shift of the three curves 
by $\Delta R\approx +0.1$ ($-0.1$).

{\em Event rates in ICECUBE.---}%
The excellent angular resolution of $0.7^\circ$ expected for ICECUBE
applies only for muon induced showers, while for $\nu_e$ and
$\nu_\tau$ events the resolution is only about 
$25^{\circ}$~\cite{Beacom:2003nh}.  According to the estimate in
Ref.~\cite{Anchordoqui:2003vc}, one expects roughly 1.5 atmospheric
$\nu_{\mu}$ background events per year  at $E>1$~TeV in a window of
$1^{\circ}\times 1^{\circ}$, i.e. $\approx$ 2.3 yr$^{-1}$ 
events in a $0.7^\circ$ radius around the Cygnus region. This number has 
to be compared with the $\approx 4\, \bar\nu_{\mu}$ signal events 
assuming $\theta_{23}=45^\circ$ and $\theta_{13}=0$. 
A $2~\sigma$ detection of the $\bar{\nu}_{\mu}$ flux is then within 1 yr 
capability of ICECUBE.
Rescaling this background number to a cone of $25^{\circ}$ opening angle, one
expects about 2900 $\nu_{\mu}$ background events and 145
background showers. Here we used the fact that the atmospheric neutrino
background has a flavor ratio of $\phi_e:\phi_\mu:\phi_\tau\approx
0.05:1:0$ in the energy range of interest, $10^{11}~{\rm eV} \alt E\alt
10^{14}$~eV~\cite{Beacom:2004jb}. 
The resulting statistical fluctuation of the background shower
number is $\sqrt{N}\approx 12$. Thus integrating one year the 
$\approx$ 16 yr$^{-1}$ rate from Cygnus one expects a $1.3~\sigma$
signal hint, or equivalently a $4.2~\sigma$ measurement in a decade.

Obviously, the poor angular resolution for $\nu_e$ and $\nu_\tau$
events is the most serious obstacle to improve this measurement.
If however a future neutrino telescope would be able to increase the shower
resolution to, say, $10^\circ$, then the same estimate would lead to a
$3.3~\sigma$ detection already in one year of data taking.
Theoretical predictions for the neutron spectrum at the source could 
also be used to optimize the detection strategy. 
To fit the anisotropy data without introducing a cutoff, the 
AGASA collaboration required in~\cite{Hayashida:1998qb} a source spectrum
with $\propto E^{-3}$ or steeper, while the spectral index of the model
of Ref.~\cite{Anchordoqui:2003vc} is 3.1. The atmospheric
neutrino flux falls with a similar slope: its spectral index 
is in the range 3--3.7, being steeper at higher energies.
Thus, if the $\bar{\nu}_e$ spectrum would be truly harder than the 
atmospheric neutrino background, the signal to background ratio could be 
improved by an increase of the threshold energy. Notice also that
experimentally, the energy spectrum of the signal events could be more 
easily measured using the shower events~\cite{Beacom:2004jb}. 

What happens to our previous estimates if we add some contamination
from ``conventional'' pion decay? If the nuclei photo-dissociation
mechanism is the correct explanation for the neutron signal, realistic
models as the one for the Cygnus region considered 
in~\cite{Anchordoqui:2003vc} would led to $\cal{O}$(10\%) flux ``pollution''. 
In this case, a shift as low as 0.01--0.02 is expected in the flux ratio $R$, 
well within the expected experimental statistical error. 
An accidental pion contamination of the same order of the expected 
signal would lead to shifts of $\approx$ +0.1 in $R$: the 
parameter estimate would then be challenging, but significant constraints
on the parameter space would be still possible, in particular when 
$\nu$-telescopes data could be combined with complementary information from 
terrestrial experiments.
Finally, we want to add a remark on the case when neutrons are
generated \emph{mainly\/} in $pp$ or $p\gamma$ collisions. Since the
normalization of the $\bar{\nu}_e$ flux from neutron decay is
based on the $\approx$ 4\% anisotropy in the CR data, the number of
events in ICECUBE from neutron decay does not 
depend on the specific generation mechanism. However, when neutrons
are produced in $pp$ or $p\gamma$ collisions, additionally a much
larger flux of neutrinos from pion decays with
$\phi_e^D:\phi_\mu^D:\phi_\tau^D\approx 1:1:1$ is expected.
Obviously, the background for the $\theta_{13}$
and $\dcp$ searches discussed here would therefore drastically
increase, while the detection of these galactic point sources by neutrino
telescopes would become much easier. A much larger flux and a flavor ratio
$\phi_\mu^D/(\phi_e^D+\phi_\tau^D)\approx 1/2$ in ICECUBE would be a
smoking gun for the dominance of the $pp$ or $p\gamma$ collision mechanism.
Although less exciting from the point of view of neutrino physics,
such a measurement would have important consequences for the
astrophysical source diagnostics as well as for CR composition
studies at $\sim 10^{18}$ eV.

{\em Summary.---}%
It has been argued that the excess of high-energy cosmic rays from the
Galactic Plane in the energy range around $10^{18}$~eV is caused by
neutron primaries generated in the photo-dissociation of heavy
nuclei. If this model is correct, then the initial flavor ratio of the
neutrino flux from the Cygnus region is
$\phi_e:\phi_\mu:\phi_\tau\approx 1:0:0$.
Thus Nature may provide in a very cheap way almost pure flavor
neutrino beams, that similarly to proposed beta-beam
factories~\cite{Zucchelli:2002} might help to deepen our knowledge of
the neutrino mixing parameters. 
In particular, we have shown that the observable ratio
$\phi^D_\mu/(\phi^D_e+\phi^D_\tau)$  of track to shower
events in a neutrino telescope depends appreciably on the 13-mixing angle
$\theta_{13}$ and the leptonic CP phase $\dcp$, thus opening a new
experimental avenue to measure these quantities.

Obviously, a better theoretical modeling of sources as well as
more experimental studies, not only in cosmic rays but also in the
photon channel, are highly desirable. Especially worthwhile would
be a confirmation of the anisotropy by the Auger
observatory~\cite{Auger} and more detailed chemical composition
studies by the Kascade-Grande experiment~\cite{KASCADE}.

\begin{acknowledgments}
We are grateful to M. Lindner, A. Mirizzi, G. Mangano, G. Raffelt, and
especially to J. Beacom for useful discussions. MK acknowledges an Emmy
Noether grant of the Deut\-sche For\-schungs\-ge\-mein\-schaft.
\end{acknowledgments}



\begin{thebibliography}{00}

\vspace*{-0.4cm}
\bibitem{review}
For a review of the current status of neutrino oscillations see e.g.
the focus issue
New J.\ Phys.\  {\bf 6} (2004)
or the Proceedings of
``21st Inter. Conf. on Neutrino Physics and Astrophysics (Neutrino 2004),'' 
Nucl.\ Phys.\ B (Proc. Suppl.) {\bf 143}, 3 (2005).


\bibitem{Apollonio:2002gd}
M.~Apollonio {\it et al.},
Eur.\ Phys.\ J.\ C {\bf 27}, 331 (2003)
[hep-ex/0301017].
	
\bibitem{sub}
M.~C.~Gonzalez-Garcia, M.~Maltoni, C.~Pena-Garay and J.~W.~F.~Valle,
Phys.\ Rev.\ D {\bf 63}, 033005 (2001)
[hep-ph/0009350];
S.~Goswami and A.~Y.~Smirnov,
hep-ph/0411359;
E.~K.~Akhmedov, A.~Dighe, P.~Lipari and A.~Y.~Smirnov,
Nucl.\ Phys.\ B {\bf 542}, 3 (1999)
[hep-ph/9808270];
J.~Bernabeu, S.~Palomares Ruiz and S.~T.~Petcov,
Nucl.\ Phys.\ B {\bf 669}, 255 (2003)
[hep-ph/0305152].
O.~L.~G.~Peres and A.~Y.~Smirnov,
Nucl.\ Phys.\ B {\bf 680}, 479 (2004)
[hep-ph/0309312].

\bibitem{f1}
K.~Anderson {\it et al.},
hep-ex/0402041.

\bibitem{Huber:2004ug}
P.~Huber {\it et al.},
Phys.\ Rev.\ D {\bf 70}, 073014 (2004)
[hep-ph/0403068].

\bibitem{BG}
P.~Bhattacharjee and N.~Gupta,
hep-ph/0501191.

\bibitem{GG}
M.~C.~Gonzalez-Garcia, F.~Halzen and M.~Maltoni,
hep-ph/0502223.


\bibitem{Beacom:2002vi}
J.~F.~Beacom {\it et al.},
Phys.\ Rev.\ Lett.\  {\bf 90}, 181301  (2003)
[hep-ph/0211305].

\bibitem{Beacom:2003zg}
J.~F.~Beacom {\it et al.},
Phys.\ Rev.\ D {\bf 69}, 017303  (2004)
[hep-ph/0309267].

\bibitem{Hooper:2004xr}
D.~Hooper, D.~Morgan and E.~Winstanley,
Phys.\ Lett.\ B {\bf 609}, 206 (2005)
[hep-ph/0410094].

\bibitem{Anchordoqui:2003vc}
L.~A.~Anchordoqui, H.~Goldberg, F.~Halzen and T.~J.~Weiler,
Phys.\ Lett.\ B {\bf 593}, 42  (2004)
[astro-ph/0311002].

\bibitem{Crocker:2004}
R.~M.~Crocker {\it et al.},
Astrophys.\ J.\  {\bf 622}, 892 (2005)
[astro-ph/0408183];
astro-ph/0411471.

\bibitem{Hayashida:1998qb}
N.~Hayashida {\it et al.}  [AGASA Collaboration],
Astropart.\ Phys.\  {\bf 10}, 303 (1999)
[astro-ph/9807045].

\bibitem{Teshima2001}
M.~Teshima {\it et al.}, in
Proc. 27th ICRC, Copernicus Gesellschaft, 2001, p.341.

\bibitem{Bird:1998nu}
D.~J.~Bird {\it et al.}  [HIRES Collaboration],
Astrophys.\ J.\  {\bf 511}, 739 (1999)
[astro-ph/9806096].

\bibitem{Bellido:2000tr}
J.~A.~Bellido, R.~W.~Clay, B.~R.~Dawson and M.~Johnston-Hollitt,
Astropart.\ Phys.\  {\bf 15}, 167 (2001)
[astro-ph/0009039].

\bibitem{Aharonian:2002ij}
F.~A.~Aharonian {\it et al.},
Astron.\ Astrophys.\  {\bf 393}, L37  (2002) 
[astro-ph/0207528].

\bibitem{Aharonian:2005}
F. Aharonian {\it et al.},
astro-ph/0501667

\bibitem{Butt:2003xc}
Y.~Butt {\it et al.},
Astrophys.\ J.\  {\bf 597}, 494 (2003)
[astro-ph/0302342].

\bibitem{Aharonian:2004wa}
F.~Aharonian {\it et al.}  [The HESS Collaboration],
astro-ph/0408145.

\bibitem{Crocker:1999yw}
R.~M.~Crocker, F.~Melia and R.~R.~Volkas,
Astrophys.\ J.\ Suppl.\  {\bf 130}, 339 (2000)
[astro-ph/9911292].


\bibitem{Berezinsky:2004wx}
V.~S.~Berezinsky, S.~I.~Grigorieva and B.~I.~Hnatyk,
Astropart.\ Phys.\  {\bf 21}, 617 (2004)
[astro-ph/0403477].


\bibitem{watson}
A.~A.~Watson,
astro-ph/0410514.

\bibitem{ss}
F.~W.~Stecker and S.~T.~Scully,
Astropart.\ Phys.\  {\bf 23}, 203 (2005)
[astro-ph/0412495].

\bibitem{ice}
A.~R.~Fazely  [The IceCube Collaboration],
astro-ph/0406125.
See also \url{http://icecube.wisc.edu}

\bibitem{Akhmedov:2004ny}
E.~K.~Akhmedov {\it et al.},
JHEP {\bf 0404}, 078 (2004)
[hep-ph/0402175].

\bibitem{Beacom:2003nh}
J.~F.~Beacom {\it et al.},
Phys.\ Rev.\ D {\bf 68}, 093005  (2003)
[hep-ph/0307025].

\bibitem{Gandhi:1998ri}
R.~Gandhi, C.~Quigg, M.~H.~Reno and I.~Sarcevic,
Phys.\ Rev.\ D {\bf 58}, 093009 (1998)
[hep-ph/9807264].

\bibitem{Beacom:2004jb}
J.~F.~Beacom and J.~Candia,
JCAP {\bf 0411}, 009 (2004)
[hep-ph/0409046].

\bibitem{Zucchelli:2002}
P.~Zucchelli,
Phys.\ Lett.\ B {\bf 532}, 166 (2002).

\bibitem{Auger}
J.~W.~Cronin,
Nucl.\ Phys.\ Proc.\ Suppl.\  {\bf 28B}, 213 (1992);
see also \url{http://www.auger.org/}

\bibitem{KASCADE}
G.~Navarra {\it et al.},
Nucl.\ Instrum.\ Meth.\ A {\bf 518}, 207 (2004);
see also \url{http://www-ik.fzk.de/KASCADE_home.html}

\end{thebibliography}
\end{document}